\documentclass[twocolumn, fleqn]{doc_class}

\usepackage[super, comma, sort&compress]{natbib}

\usepackage{newtxtext, newtxmath}

\usepackage[T1]{fontenc}
\usepackage[utf8]{inputenc}

\usepackage{ulem}  %
\newcommand{\redsout}[1]{}

\usepackage{ae, aecompl}
\usepackage{graphicx}
\usepackage[dvipsnames, hyperref]{xcolor}
\usepackage{xspace}
\usepackage{soul}
\usepackage{booktabs}
\usepackage{adjustbox}
\usepackage{comment}

\usepackage{amsmath}
\usepackage{commath}
\usepackage[separate-uncertainty,bracket-ambiguous-numbers=false,separate-uncertainty-units=single]{siunitx}
\usepackage{nth}

\usepackage[capitalise, noabbrev]{cleveref}
\usepackage{enumitem}
\usepackage{csvsimple}
\usepackage{multirow}
\usepackage{array}
\newcolumntype{P}[1]{>{\centering\arraybackslash}p{#1}}

\crefname{figure}{Fig.}{Figs.}
\crefname{equation}{equation}{equations}

\usepackage[leftmargin=0pt,rightmargin=0pt,skipabove=2pt,skipbelow=2pt,innerleftmargin=5pt,innerrightmargin=5pt,innertopmargin=5pt,innerbottommargin=5pt,linewidth=0pt,innerlinewidth=0pt,middlelinewidth=0pt,outerlinewidth=0pt,roundcorner=0pt]{mdframed}

\DeclareSIUnit{\MSun}{\ensuremath{M_\odot}}
\DeclareSIUnit{\Msun}{\ensuremath{M_\odot}}
\DeclareSIUnit[quantity-product = ]\percent{\char`\%}

\newcommand{\LCDM}{$\Lambda$CDM\xspace}
\usepackage{physics}
\newcommand*{\borg}{\textsc{borg}\xspace}
\newcommand{\citem}{\cite}
\newcommand{\citemet}{\cite}

\newcommand{\orcidsymb}[2]{#1\href{http://orcid.org/#2}{\adjustbox{trim={-.15\width} {0\height} {-.15\width} {0\height},clip}{\includegraphics[height=10pt]{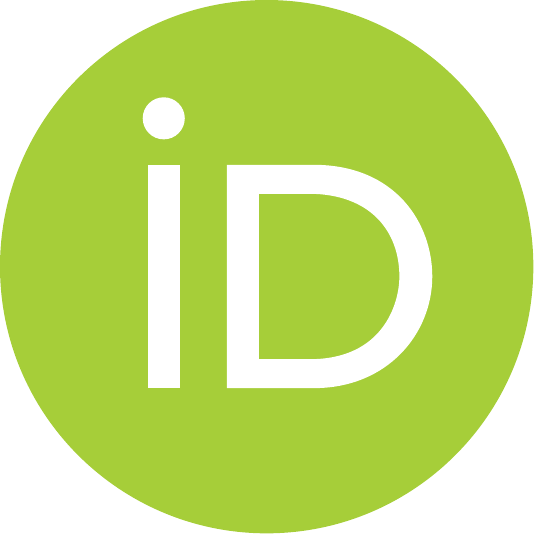}}}}

\hypersetup{
    unicode=true,
    final=true,
    plainpages=false,
    pdfstartview=FitV,
    pdftoolbar=false,
    pdfmenubar=true,
    bookmarksopen=true,
    bookmarksnumbered=true,
    breaklinks=true,
    linktocpage,
    colorlinks=true,
    linkcolor=Blue,
    urlcolor=Blue,
    citecolor=Blue,
    anchorcolor=green
}
\AtBeginDocument{
    \hypersetup{
        pdftitle={The mass distribution around the Local Group}, %
        pdfauthor={E. Wempe, S.D.M. White et al.},
        pdfsubject={}%
    }
}

\newcommand\add[1]{#1}

\usepackage{paralist}
\makeatletter
    \def\@biblabel#1{\@ifnotempty{#1}{#1}}
    \def\NAT@anchor#1#2{
        \hfilneg\hyper@natanchorstart{#1\@extra@b@citeb}
        #2.
        \hyper@natanchorend
    }
\makeatother
\DeclareRobustCommand{\VAN}[3]{#2}
\let\VANthebibliography\thebibliography
\def\thebibliography{\DeclareRobustCommand{\VAN}[3]{##3}\VANthebibliography}

\title[The mass distribution in and around the Local Group]{The mass distribution in and around the Local Group}

\author[]{{\orcidsymb{Ewoud Wempe}{0000-0001-8232-4188}$^{{\hyperlink{inst:Kapteyn}{1}},\star}$, \orcidsymb{Simon D.M. White}{0000-0002-1061-6154}$^{\hyperlink{inst:MPA}{2}}$, \orcidsymb{Amina Helmi}{0000-0003-3937-7641}$^{\hyperlink{inst:Kapteyn}{1}}$, \orcidsymb{Guilhem Lavaux}{0000-0003-0143-8891}$^{\hyperlink{inst:CNRS}{3}}$, \orcidsymb{Jens Jasche}{0000-0002-4677-5843}$^{\hyperlink{inst:SU}{4}}$}
\\
\\
\hypertarget{inst:Kapteyn}$^{1}$ Kapteyn Astronomical Institute, University of Groningen, P.O Box 800, 9700 AV Groningen, The Netherlands
\\
\hypertarget{inst:MPA}$^{2}$ Max-Planck-Institut für Astrophysik, Karl-Schwarzschild-Straße 1, 85748 Garching, Germany
\\
\hypertarget{inst:CNRS}$^{3}$ CNRS \& Sorbonne Université, UMR 7095, Institut d’Astrophysique de Paris, 98 bis boulevard Arago, F-75014 Paris, France
\\
\hypertarget{inst:SU}$^{4}$ The Oskar Klein Centre, Department of Physics, Stockholm University, Albanova University Center, SE 106 91 Stockholm, Sweden\\
{$^\star$ \normalfont E-mail: \href{mailto:ewoudwempe@gmail.com}{ewoudwempe@gmail.com}}
\vspace{-2em}
}

\pubyear{2025}

\begin{document}
\raggedbottom
\maketitle

\begin{abstract}
    \begin{mdframed}[backgroundcolor=black!5]
Our Galaxy, Andromeda and their companion dwarf galaxies form the Local Group. Most of the mass in and around it is believed to be dark matter rather than gas or stars, so its distribution must be inferred from the effect of gravity on the motion of visible objects. Modelling efforts have long struggled to reproduce the quiet Hubble flow around the Local Group, as they require unrealistically little mass beyond the haloes of the two main galaxies. Here we revisit this using \LCDM simulations of Local Group analogues with initial conditions constrained to match the observed dynamics of the two main haloes and the surrounding flow. The observations are reconcilable within \LCDM, but only if mass is strongly concentrated in a plane out to \SI{10}{Mpc}, with the surface density rising away from the Local Group and with deep voids above and below. This configuration, dynamically inferred, mirrors known structures in the nearby galaxy distribution. The resulting Hubble flow is quiet yet strongly anisotropic, a fact obscured by the paucity of tracers at high supergalactic latitude. This flattened geometry reconciles the dynamical mass estimates of the Local Group with the surrounding velocity field, thus demonstrating full consistency within the standard cosmological model.

    \end{mdframed}

\end{abstract}
\nokeywords

\noindent The earliest quantitative inference of the local mass distribution treated the main bodies of the Local Group---the Milky Way (MW) and Andromeda (M31)---as a Keplerian two-body system. The so-called timing argument \citem{kahnIntergalacticMatterGalaxy1959} assumes that the two galaxies are point masses on a radial orbit, starting together at the Big Bang. This relates their total mass $M_{\rm TA}$ to the time since the Big Bang $t_0$, their current separation $r$ and their velocity of approach $v$. On inserting estimates of these quantities, a total mass much larger than that of the visible stars and gas was obtained \citem{kahnIntergalacticMatterGalaxy1959} for the first time in 1959.

Since then, measurements of the three observables used by the timing argument have improved substantially\citem{vandermarelM31VelocityVector2012}. The principal uncertainty in the resulting mass estimate is its interpretation, as a pair of point masses is undoubtedly a very poor representation of the extended mass distribution surrounding the two big galaxies. Isolated galaxy pairs in large-scale simulations of cosmic structure formation have previously been used\citem{liMassesLocalGroup2008} to calibrate the relation between $M_{\rm TA}$ and the sum of the virial masses of the two systems, finding $\sum M_\text{200,c}/M_{\rm TA}$ to be almost unity on average but with a broad 5\% to 95\% range, 0.33 to 2.0. With current observational data\citem{vandermarelM31VelocityVector2012}, this implies that $\sum M_\text{200,c}$ values below about \SI{2e12}{\Msun} are strongly disfavoured.

One can also use the motions of neighbouring objects to constrain the mass of the Local Group. If these are sufficiently distant, the MW--M31 pair can be approximated as a single object at their barycentre and a spherical infall model can be fitted to the motions of the nearby galaxies, thus reducing their equations of motion to the timing argument form, except that $M_{\rm TA}$ is replaced for each tracer by the total mass $M(r)$ within its distance $r$ from the Local Group barycentre. Together with $t_0$ and an assumed cosmological constant, this leads to a velocity versus distance relation $v(r)$ that can be compared with the observed recession velocities of the tracer galaxies. In the absence of significant excess mass in the Local Group and its surroundings, the velocity--distance relation would simply follow the Hubble law, $v(r)=Hr$, where $H$ is the Hubble constant. However, the gravitational pull of the Local Group reduces velocities relative to this expectation. The infall velocities depend on the assumed $M(r)$, which must increase with $r$ and must exceed the mass of the Local Group at all external radii. An extreme assumption is that, at all radii, $M(r)$ is dominated by the haloes of M31 and the MW and so can be approximated as a constant. Fitting such a model to the motions of nearby galaxies is clearly expected to produce an upper limit on the true value of $\sum M_\text{200,c}$.

Under this extreme assumption, the timing argument for the MW--M31 pair can be combined with an estimate of the distance to the zero-velocity surface of the Local Group (the ``turnround'' radius, $r_0\approx \SI{1}{Mpc}$, such that $v(r_0)=0$) to infer both the mass of the Local Group and the age of the Universe\citem{lynden-bellDynamicalAgeLocal1981}. Extensions of this argument \citem{giraudPerturbationNearbyExtragalactic1986,sandageRedshiftDistanceRelationIX1986}
out to several megaparsecs have inferred upper limits of \SI{\sim3e12}{\Msun} and a residual scatter around a spherically symmetric infall of \SI{50}{km.s^{-1}} to \SI{60}{km.s^{-1}}.
Subsequent use of an updated and more local catalogue ($r< \SI{3}{Mpc}$),
in conjunction with other constraints, has yielded estimates of the Local Group mass ranging \citem{karachentsevVeryLocalHubble2002,penarrubiaDynamicalModelLocal2014} from just \SI{1.3(3)e12}{\Msun}
to \SI{2.3(7)e12}{\Msun} and an even quieter flow, with a dispersion in the infall velocity of only \SI{\sim30}{km.s^{-1}}. This mass should again be interpreted as an upper limit on $\sum M_\text{200,c}$ because the method assumes this to be the only significant mass in the region. It is striking that all these upper limits are close to the lower limit on $\sum M_\text{200,c}$ given by the timing argument alone and discussed above.

It is clearly unrealistic to assume that the only mass within \SI{3}{Mpc} or \SI{5}{Mpc}, or even within the zero-velocity surface of the Local Group, is that within the virial radii of M31 and the MW. On the other hand, including extra material within and around the Local Group will only increase the infall velocity, at least in the spherical approximation, and hence also increase the tension with timing argument estimates of $\sum M_\text{200,c}$. To understand whether this tension reflects an issue with the observational data, a misunderstanding of cosmic structure formation in the standard paradigm or even a failure of that paradigm, a more principled approach is needed. We have recently provided such an approach \citem{wempeConstrainedCosmologicalSimulations2024} by adapting the Bayesian Origin Reconstruction from Galaxies (BORG) formalism \citem{jascheBayesianPhysicalReconstruction2013,jaschePhysicalBayesianModelling2019} to generate a statistically representative ensemble of $\Lambda$ cold dark matter (\LCDM) universes that contain realistic Local Group analogues. Briefly, a hierarchical Bayesian inference is run on the primordial density field while assuming a \LCDM prior and a likelihood based on present-day constraints. Haloes were required to form at the observed positions of M31 and the MW with masses and relative velocity consistent with the motions of observed tracers around each galaxy. In addition, the recession velocity was required to be consistent with observations at the positions of 31 isolated external galaxies out to \SI{4}{Mpc}. These galaxies trace the local Hubble flow, within some scatter, estimated\citem{penarrubiaDynamicalModelLocal2014} to be \SI{35}{km.s^{-1}}.

This programme successfully generated a representative ensemble of realizations that simultaneously reproduce all the observations within their estimated uncertainties. To enhance the resolution beyond that of the BORG realizations, we here use a set of 169 Gadget-4 resimulations of the selected sample of quasi-independent initial conditions\citem{wempeEffectEnvironmentMass2025} (Methods).
These simulations allow us to assess if and how the weakly perturbed Hubble flow surrounding the Local Group can be reconciled with the relatively high value of $\sum M_\text{200,c}$ implied by the timing argument.

\begin{figure}
    \centering
    \includegraphics[width=\linewidth]{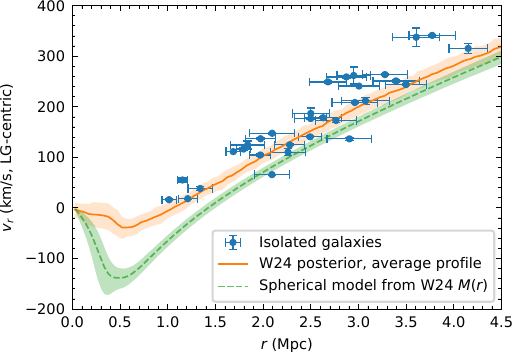}
    \caption{Hubble diagram for the isolated galaxy sample used to trace the velocity field beyond the Local Group. Error bars indicate $1\sigma$ measurement uncertainties on the observed distances and velocities. These velocities, together with the masses, relative positions and motions of the M31 and MW haloes, were used to constrain the initial conditions of a representative ensemble of 169 simulations of the Local Group (ref.~\protect\citenum{wempeConstrainedCosmologicalSimulations2024}; labelled W24 in the legend) within the \LCDM paradigm.
    The orange line shows the mean velocity--distance relation for this ensemble. The green line shows the relation obtained from a spherical infall model based on the mean mass profile $M(r)$ of this same ensemble. Shaded regions represent the 16th to 84th percentiles of the simulation-to-simulation scatter (the posterior scatter). Clearly, the two curves are not consistent.}
\label{fig:hubblediagram}
\end{figure}

\section*{Results}
Figure~\ref{fig:hubblediagram} is a Hubble diagram. The points in blue show the 31 isolated galaxies used to constrain the velocity field. These are plotted in a Local Group reference frame, which we centre at a distance of \SI{495}{kpc} along the line connecting the MW and M31---approximately 2/3 of the way from the MW to M31---assuming that the MW is falling towards this point with $\sim$2/3 of the relative radial velocity of the two galaxies (as inferred in previous work\citem{wempeConstrainedCosmologicalSimulations2024,penarrubiaDynamicalModelLocal2014}).

\begin{figure*}
    \centering
    \includegraphics[width=\linewidth]{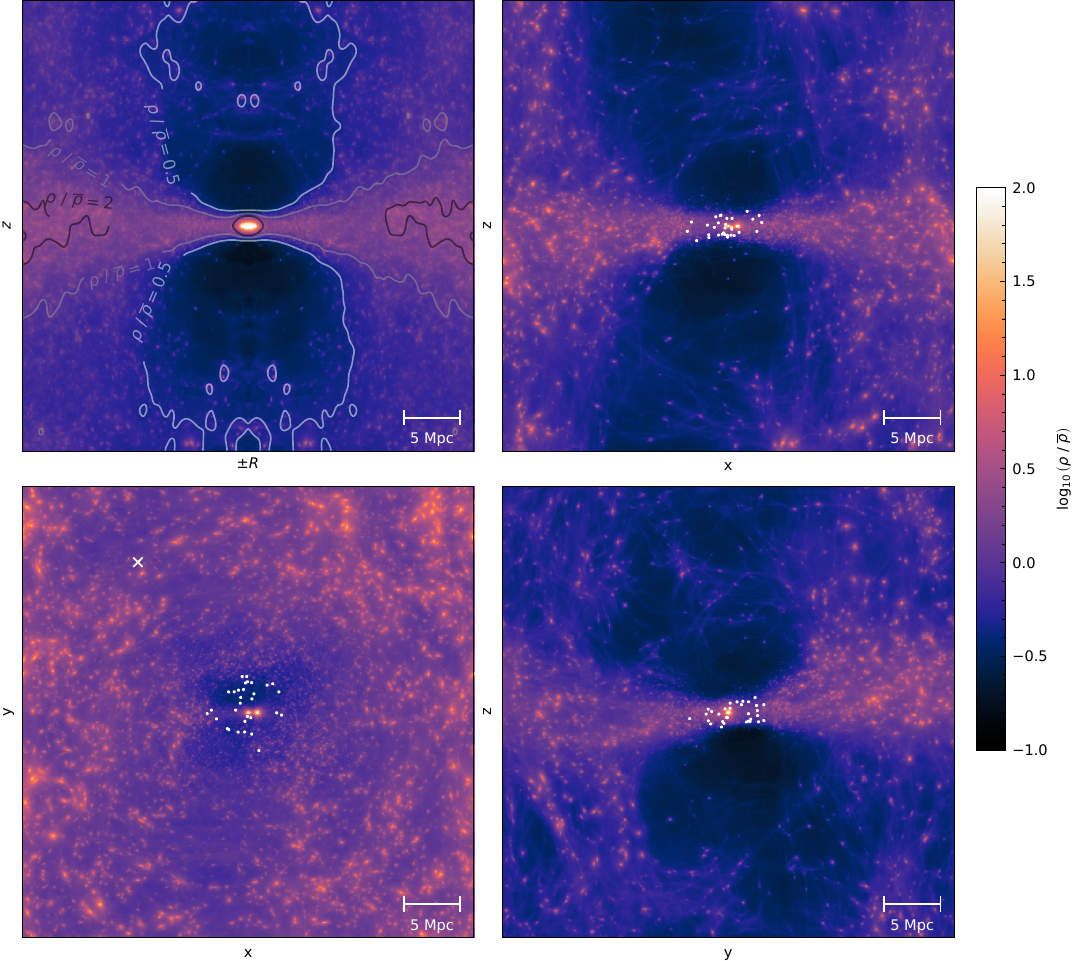}
    \caption{Various projections of the posterior mean density of the constrained simulation ensemble, normalized by the cosmic mean density, $\overline{\rho}$. The coordinate system is centred on the MW, such that $x$ is along the MW--M31 axis, $y$ is towards increasing declination and $z$ is towards increasing right ascension at the position of M31. Upper left: the density in cylindrical coordinates azimuthally averaged around the $z$ axis with a few representative contours (these were drawn after smoothing with a \SI{40}{kpc} Gaussian filter). Lower left: a projection along the $z$ axis of a slice of depth \SI{8}{Mpc} ($\abs{z} < \SI{4}{Mpc}$). The right panels show the remaining two orthogonal projections, each corresponding to a central slice of depth \SI{8}{Mpc}. White dots indicate the locations of the isolated galaxies used as flow tracers. The white cross indicates the location of the Virgo cluster. No clear overdensity is seen in this direction, and we explicitly checked for any signal with respect to azimuthal angle $\phi - \phi_\text{Virgo}$, finding nothing. This is consistent with the fact that the mean infall to Virgo is removed in the velocity frame of our simulations and that tidal effects due to Virgo should be relatively small over the $\SI{<4}{Mpc}$ region where we have included constraints.
    The density visualizations use a Lagrangian sheet interpolation scheme (Methods).}
    \label{fig:sheet}
\end{figure*}

\interfootnotelinepenalty=10000
The mass-weighted radial velocity profile of our ensemble of Local Group analogues, \begin{equation}
    \overline{v_r}(r) = \frac{1}{M_\text{shell}}\sum_{\text{particle $i$ in shell}}{m_i\frac{\vb*v_i \vdot(\vb*r_i - \vb*r_\text{LG})}{\norm{\vb*r_i - \vb*r_\text{LG}}}},
\end{equation}
where the sum runs over radial shells of width \SI{50}{kpc} about the Local Group barycentre $r_\text{LG}$ with $M_\text{shell}$ the total mass in each shell, and $m_i$, $r_i$ and $v_i$ the particle masses, positions and velocities computed in radial bins of \SI{50}{kpc} about this Local Group centre, is shown by the orange line in \cref{fig:hubblediagram}. The velocity--radius relation implied by a spherical infall model (Methods) %
is shown in green.
Although the observational data match the actual mass-weighted velocity profile of the simulations reasonably well, the spherical model fails badly; the predicted velocities lie significantly below the data at all radii. \add{The posterior uncertainty in the enclosed mass is insufficient to explain this; the spherical model is $5.4\sigma$ below the ensemble mean flow at \SI{1}{Mpc} and $3.5\sigma$ below it at \SI{2}{Mpc}.} As discussed earlier, matching the observations with a spherical model implies very little mass within \SI{4}{Mpc} beyond that in the haloes of two main galaxies\add{ --- the sum of the two halo masses alone is at least \SI{2e12}{\Msun}, and in our inference, we find $\sum M_\text{200,c}=\SI{3.3(6)e12}{\Msun}$, yet a spherical model with a total mass of just \SI{\sim3e12}{\Msun} within \SI{1}{Mpc} and no external mass at all fits the tracer data well out to \SI{4}{Mpc} (ref.~\citenum{wempeConstrainedCosmologicalSimulations2024})}. This is unrealistic in reasonable models of structure formation; in our ensemble of Local Group analogues, the mean mass within \SI{4}{Mpc} is \add{$4.2\times \sum M_\text{200,c}$}. \add{The green curve illustrates the consequence of this conflict: including this external mass in a spherical model invalidates the fit to the data.}

The reason that our simulations fit the velocity field whereas a spherical model does not is that we infer a mass distribution that is \add{not spherically symmetric but, rather, is sheet-like.
In a spherically symmetric system, the net force at each radius is determined solely by the enclosed mass. This is not the case in a strongly flattened system, however. Mass located at larger cylindrical radii but near the plane exerts an outwards gravitational pull that partially offsets the inwards force experienced by the tracers, hence reducing their present-day infall velocities or, equivalently, increasing their recession velocities.}

This \add{sheet-like configuration} is especially clear from our inferred posterior mean density field (the average over the ensemble of all constrained simulations). In \cref{fig:sheet}, several projections of this posterior mean are shown in simulation coordinates.
The top left panel shows a cylindrical view averaged azimuthally around the $z$ axis. Clearly, there is a strong concentration of mass towards the equatorial plane of this coordinate system. We also show central slices of depth \SI{8}{Mpc} of three Cartesian projections: the bottom left panel shows a view projected along the $z$ axis, and the right two panels show the two orthogonal projections.
One can see that the mass around the MW and M31 is, on average, concentrated in a sheet that is aligned with the $x$--$y$ plane.
Interestingly, in this coordinate system, $z$ is misaligned by only \SI{12}{deg} from the north pole of the Local Sheet of galaxies, which is itself closely aligned with the Supergalactic Plane \citem{tullyOurPeculiarMotion2008,mccallCouncilGiants2014}. We have inferred this mass sheet by requiring consistency between Local Group dynamics and the surrounding Hubble flow, yet it closely reflects the structure traced by luminous galaxies in the nearby Universe.

\begin{figure}
    \centering
    \includegraphics[width=\linewidth]{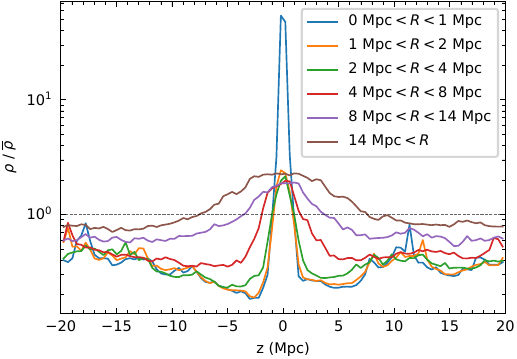}
    \caption{Azimuthally averaged density as a function of height $z$ within thick cylindrical shells centred at the MW--M31 barycentre. Here $R = \sqrt{x^2 + y^2}$ and the $z$-axis is the simulation axis most nearly perpendicular to the Local Sheet.}
    \label{fig:densityrings}
\end{figure}

The overdensity profile of the sheet is illustrated more quantitatively in \cref{fig:densityrings}.
We find over the entire simulation region that the azimuthally averaged density on the midplane of the sheet is about twice the cosmological mean, $\overline{\rho}$\add{, with the core of the sheet (defined as a cylindrical shell with $1<R<\SI{4}{Mpc}$ and $|z|<\SI{0.5}{Mpc}$) having vertically averaged density $\rho/\overline{\rho}=\num{2.03(0.78:0.76)}$. (Here and below quantities given in this form refer to the mean and the 16 and 84\% points of the distribution over the individual simulations of our ensemble.)
The thickness of the central overdensity spike, defined as the full vertical width over which the smoothed density (averaged within $R < \SI{4}{Mpc}$ and smoothed in $z$ by a \SI{40}{kpc} kernel) remains above the cosmic mean, is \SI{1.64(0.45:0.42)}{Mpc}.}
Furthermore, the vertical scale height of this overdensity increases with $R$, so that the projected surface density of the plane increases with distance from the Local Group, as is clearly visible in the lower left panel of \cref{fig:sheet}. In addition, there are strong underdensities above and below the plane, \add{with $\rho/\overline{\rho}=\num{0.26(0.14:0.12)}$ (for $R<4$~\si{Mpc}, $-8<z<-4$~\si{Mpc}) in the direction of the Local Void\citem{tullyAtlasNearbyGalaxies1987}, and \num{0.31(0.11:0.16)} (for $R<4$~\si{Mpc}, $4<z<8$~\si{Mpc}) towards the local mini-void\citem{karachentsevVeryLocalHubble2002}.}

Although we see this sheet unambiguously in the posterior mean field, it is also present in the individual realizations of the ensemble.
If for each simulation we compute the eigenvalues and principal axis directions of the inertia tensor for particles with distances $\SI{2}{Mpc}<r<\SI{4}{Mpc}$ 
(or alternatively $\SI{4}{Mpc}<r<\SI{8}{Mpc}$) we find the angle between the shortest axis and the $z$-direction 
to be $\SI{14(6:8)}{deg}$ 
($\SI{16(8:10)}{deg}$); the average direction (the direction of the average of the minor axis unit vectors) is just $\SI{5}{deg}$ ($\SI{6}{deg}$) from the $z$-axis. For comparison, the angle between the shortest axis and the observed Local Sheet north pole is \SI{18(7:8)}{deg} (\SI{20(9:10)}{deg}), with the average direction being \SI{13}{deg} (\SI{15}{deg}) from the pole. 

Individual realizations are also quite strongly flattened with minor-to-major axis ratios $c/a = \num{0.24(0.09:0.08)}$ (\num{0.30(0.10:0.11)}), intermediate-to-major axis ratios $b/a = \num{0.68(0.16:0.16)}$ (\num{0.72(0.15:0.14)}) and minor-to-intermediate axis ratios $c/b = \num{0.36(0.09:0.09)}$ (\num{0.41(0.12:0.12)}), measured from the mass-weighted inertia tensor in the \SIrange{2}{4}{Mpc} (\SIrange{4}{8}{Mpc}) shell.
$c/b$ is substantially smaller than $b/a$, showing that the mass distribution around the Local Group is consistently inferred to be sheet-like rather than filamentary. \add{An approximately spherical geometry around the Local Group is decisively rejected; across our 169 posterior samples, the {\it maximum} value of $c/a$ is just \num{0.45} in the 2--4~\si{Mpc} shell, whereas a spherical distribution would have $c/a\sim 1$.}
The configuration we have found bears a striking resemblance to the structure observed in the spatial distribution of nearby galaxies, specifically the Local Sheet\citem{peeblesRadialTransverseVelocities2001,tullyOurPeculiarMotion2008,peeblesNearbyGalaxiesPointers2010}, the ``Council of Giants'' \citem{mccallCouncilGiants2014} and the local voids above and below the sheet (the Local Void \citem{tullyAtlasNearbyGalaxies1987,tullyOurPeculiarMotion2008,tullyCosmicflows32016}, and a smaller local mini-void \citem{karachentsevVeryLocalHubble2002}). \redsout{, as traced by the positions of nearby galaxies.} Our measured $c/a$ and $b/a$ ratios are like those found\citem{neuzilSheetGiantsUnusual2020} for galaxy positions within \SI{8}{Mpc}, which yield
$c/a=0.16$ and $b/a=0.79$.
Our analysis is, thus, a dynamical demonstration that \add{in our local neighbourhood} light approximately traces mass also on these scales.

\begin{figure}
    \centering
    \includegraphics[width=\linewidth]{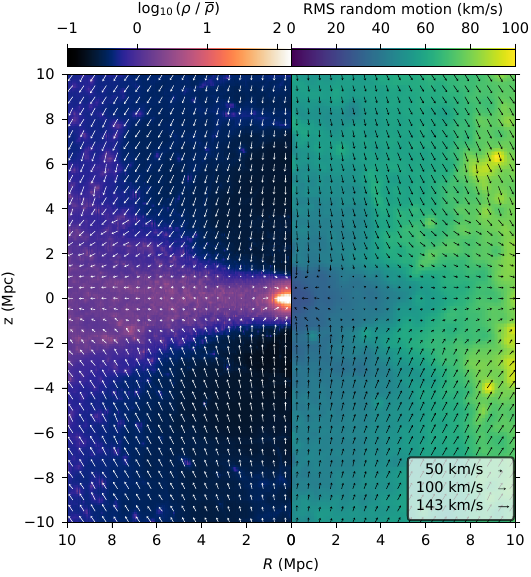}
    \caption{The peculiar velocity field in cylindrical coordinates. Each arrow shows the mean velocity in a $\Delta R = \Delta z = \SI{0.5}{Mpc}$ bin. The background colouring in the left half shows the posterior mean density field (as in \cref{fig:sheet}), while that on the right shows the \textit{r.m.s.} scatter of the individual \SI{500}{kpc} smoothed posterior velocity fields about their azimuthally averaged ensemble mean.}
    \label{fig:velocityfield}
\end{figure}
\cref{fig:velocityfield} shows the average peculiar velocity field. Particle momenta and masses from each realization are binned in cylindrical coordinates with $\Delta R = \Delta z = \SI{0.5}{Mpc}$. The stacked momentum and mass fields are then divided to obtain the mean velocity in each bin. The arrows in \cref{fig:velocityfield} show the resulting field.
This ensemble mean flow is predominantly towards the sheet from above and below.
Within the sheet, infall towards the Local Group is evident only \add{within a radius of $R=\SI{2.56(0.31:0.33)}{Mpc}$} (where the \SI{500}{kpc}-smoothed, azimuthally averaged $v_R$ changes sign) and infall velocities are relatively low; at larger distances, peculiar velocities in the sheet are actually directed {\it away} from the Local Group. Inflow velocities are much larger towards the poles and approach or exceed \SI{100}{km.s^{-1}}, even out to \SI{8}{Mpc}. This strong anisotropy is not visible in our set of flow tracers because they all lie at low latitude in supergalactic coordinates\add{, as shown explicitly in \cref{fig:sheet}.}

The background colour scale on the left side of \cref{fig:velocityfield} repeats the posterior mean overdensity already shown in \cref{fig:sheet}. That on the right side indicates the \textit{r.m.s.} deviation of individual \SI{500}{kpc} smoothed flow fields from their stacked and azimuthally averaged ensemble mean.
These random motions are quite small: in a cylinder with $R<\SI{5}{Mpc}$ and $\abs{z}<\SI{5}{Mpc}$, the overall mass-weighted \textit{r.m.s.} is only \SI{33}{km.s^{-1}} (in one dimension), whereas for \SI{10}{Mpc} limits in both $R$ and $z$, it is \SI{58}{km.s^{-1}}. Within the sheet where most of our tracers lie ($R<\SI{4}{Mpc}$ and $\abs{z}<\SI{1}{Mpc}$), it is only \add{\SI{22}{km.s^{-1}}}. This shows that, although anisotropic, the overall peculiar velocity field is predicted to be very cold, even colder than the observational estimates in the papers quoted earlier. \add{This also reflects that the posterior scatter is smallest in the region where there are the most constraints.}

\section*{Discussion}
We have shown that in the \LCDM paradigm, the observed quiet local Hubble flow can be consistent with the halo masses implied for M31 and the MW by the timing argument and by internal tracer velocities only if the mass distribution is strongly concentrated in a sheet out to at least \SI{10}{Mpc}, with substantially underdense regions both above and below this Supergalactic Plane.
The surface density of the plane is lower near the Local Group than at distances of \SI{5}{Mpc} to \SI{10}{Mpc}. In a sheet-like geometry, the velocity--distance relation depends not only on the enclosed mass, as in the spherical case, but also on the mass at larger distances; the increased mass surface density at large $R$ is responsible for the reversal of infall velocities in the sheet at $R>\SI{2.5}{Mpc}$.

\add{We have checked that the inferred sheet-like mass distribution is a genuine consequence of tracer kinematics, rather than an artefact of their anisotropic spatial distribution, by conducting a mock inference test. In this test, we replaced the observed tracer velocities with predictions from a spherical model consistent with our inferred $M(r)$, while leaving the tracer positions unchanged. As detailed in Extended Data \cref{fig:hubblediagramwmock,fig:sheetmock,fig:densityringsmock}, this test shows that our inference scheme can correctly reproduce a spherical mass profile, even when the tracer distribution is non-spherical (Extended Data \cref{fig:hubblediagramwmock}). Furthermore, the resulting posterior mean density field for the mock inference shows no preferred planar structure (Extended Data \cref{fig:sheetmock}), despite the flattened sky distribution of the tracer galaxies. This demonstrates that our flattened mass distribution is driven by the observed peculiar velocities, rather than being an imprint of their spatial distribution.}

One limitation of the simulations used here \citem{wempeConstrainedCosmologicalSimulations2024} may be the relatively small region simulated (a box \SI{40}{Mpc} on a side).
The periodic boundary conditions produce anisotropic forces on scales of \SI{10}{Mpc} or more, and they may bias the structure to align with the coordinate axes, perhaps explaining why the mass sheet we find aligns somewhat better with the $x$--$y$ plane of our simulation boxes than with the observed Local Sheet. Additionally, the average density over the entire simulated region is required to be the cosmic mean, so the substantial in-plane overdensity that is seen, somewhat surprisingly, even at the largest $R$ values in \cref{fig:densityrings}, must be compensated for by a mean underdensity out of the plane.
\add{However, these limitations primarily impact the structure on scales larger than \SI{\sim10}{Mpc} and, thus, do not alter our main conclusions, which are drawn on smaller scales where the forces remain largely isotropic.}
The simulations are, clearly, too small to properly represent the structure of the Virgo cluster and the Local Supercluster, but it is interesting that planar structure is required on scales substantially larger than the region where we have posed constraints.
The inclusion of other constraints on larger scales, as in earlier BORG applications\citem{jaschePhysicalBayesianModelling2019}, should enable the whole of the Local Supercluster to be represented realistically in addition to the Local Group and its immediate environment.

\add{Our inference allows for a systematic velocity offset between the density centre of the MW and the centre of mass of its halo via a prior with an \textit{r.m.s.} of \SI{17}{km.s^{-1}}, whereas recent measurements of the reflex motion of the inner MW induced by the Large Magellanic Cloud report values of \SI{\sim40}{km.s^{-1}} with respect to distant halo tracers \citem{yaaqibRadialVariationLMCinduced2024,chandraAllskyKinematicsDistant2025}. The effective reflex velocity relevant to our constraint is much lower because our prior applies to the mass-weighted velocity of the halo averaged with a \SI{100}{kpc} Gaussian filter, which substantially downweights the outer halo. In fact, if we convolve the observed velocity profile\cite{chandraAllskyKinematicsDistant2025} with our MW mass model and filter, we find the effective reflex velocity to be only \SI{18.8(1.2)}{km.s^{-1}}, a value fully consistent with our adopted prior. Therefore, the dynamical impact of the Large Magellanic Cloud is properly accounted for in our framework.}

\add{Revisiting the tensions outlined in the introduction, our work demonstrates that the discrepancy is not due to issues with the observational data or a failure of the \LCDM paradigm but can be resolved by the flattened geometry of the mass distribution around the Local Group.}
\add{A key, testable prediction of our work is the highly anisotropic nature of the local velocity field, with strong infall towards the Local Sheet predicted from the underdense polar regions. This flow remains observationally unconstrained in our immediate vicinity because of the lack of known high-latitude tracers within \SI{5}{Mpc}. Intriguingly, isolated high-latitude galaxies have been observed at slightly larger distances\citem{karachentsevLocalGalaxyFlows2003,tullyOurPeculiarMotion2008} with infall velocities of several hundred kilometres per second. The discovery of new and closer isolated dwarfs at high latitude could provide a decisive test of the configuration we have inferred.}

\clearpage
\makeatletter
\renewcommand{\fnum@figure}{Extended Data Fig. \thefigure}
\renewcommand{\fnum@table}{Extended Data Table \thetable}
\makeatother
\setcounter{figure}{0}
\setcounter{table}{0}
\setcounter{footnote}{0}

\section*{Methods}

\subsection*{Constrained Local Group simulations}
The methodology underlying the simulations presented here is detailed in previous works\citem{wempeConstrainedCosmologicalSimulations2024,wempeEffectEnvironmentMass2025}; here we summarize the main elements relevant to this study. The Local Group ensemble is constructed in two steps: Bayesian inference of the initial conditions, followed by higher resolution resimulations of these initial conditions.

We inferred the cosmological initial conditions from which the Local Group formed using the Bayesian Origin Reconstruction from Galaxies (\textsc{borg}) framework\citem{jascheBayesianPhysicalReconstruction2013,jaschePhysicalBayesianModelling2019}. A Hamiltonian Monte Carlo (HMC) sampler was used to explore the posterior distribution of initial-condition fields in a \SI{40}{Mpc} co-moving box augmented by a higher resolution zoom region. The initial conditions follow a $\Lambda$CDM prior with \textit{Planck} 2018 parameters\citemet{planckcollaborationPlanck2018Results2020}, and a likelihood that imposes constraints on the final state of the simulation at $z=0$.

The likelihood constrained (1) the MW and M31 masses (defined as the mass within a \SI{100}{kpc} Gaussian filter) to agree with observational estimates within their uncertainties, (2) the positions of the MW and M31 haloes to match the observed positions, with a tolerance of \SI{30}{kpc} in each direction, (3) the relative three-dimensional velocity vector of the MW--M31 pair to match the observed value within its uncertainties and (4) the peculiar velocities of 31 galaxies within \SI{\sim4}{Mpc}, selected to have low distance uncertainties and to be isolated enough to trace the large-scale flow, to match the simulated velocity field at their locations within some scatter (taken to be \SI{35}{km.s^{-1}}; ref.~\citenum{penarrubiaDynamicalModelLocal2014}). This yielded a posterior ensemble of initial conditions that evolved into a Local Group analogue consistent with the surrounding Hubble flow constraints.

\add{From the converged HMC chains, we selected 169 realizations of the initial density field. We increased the resolution of these initial conditions by augmenting the \textsc{borg} fields with small-scale fluctuations drawn from the $\Lambda$CDM prior below the original Nyquist frequency and by increasing the number of particles in the zoom region. These upgraded initial conditions were evolved from $z=63$ to $z=0$ using \textsc{Gadget-4}\citemet{springelSimulatingCosmicStructure2021}. The high-resolution particles have a mass of \SI{1.88e7}{\Msun} and a co-moving force softening length of \SI{1.5}{kpc.h^{-1}}, and the low-resolution particles have masses of \SI{9.65e9}{\Msun} and softening lengths of \SI{24}{kpc.h^{-1}}. Further numerical details are provided in the relevant paper\citem{wempeEffectEnvironmentMass2025}.}

\subsection*{Spherical collapse model}
In this work we compare the dynamics in constrained simulations to the dynamics of the spherical collapse model that has traditionally been used to model the Hubble flow surrounding the Local Group.
In the spherical collapse model, the quantity $E(M) = \frac{1}{2}\dot{r}^2 - G M / r - \frac{1}{2}H_0^2\Omega_{\Lambda,0}r^2$, analogous to the specific orbital energy, is conserved until shells cross, where $G$ is the gravitational constant, and $H_0$ and $\Omega_{\Lambda,0}$ are the present-day Hubble constant and dark-energy density parameter. Here the physical radius $r$ is taken to be a function of the enclosed mass $M$ and of time. For a shell presently at radius $r_0$ and enclosing mass $M$, $E$ is found numerically by integrating $\dot{r}$ at constant $E$ and $M$ subject to the initial condition $r=0$ at $t=0$ and requiring the solution to pass through $r_0$ at $t=t_0=\SI{13.8}{Gyr}$, the current age of the Universe. The solution then also gives $\dot{r}(t_0)$, the current shell velocity, with its sign switching at apocentre.

\section*{Data availability}
The full datasets generated in this study are too large for direct public hosting but are available on request from the corresponding author.
The intermediate data needed to reproduce all results and figures are available within Supplementary Code 1 provided with this publication on the journal website.

\section*{Code availability}
The code and analysis scripts required to reproduce all results and figures in this article are available as Supplementary Code 1 with this publication on the journal website.

\begingroup
\raggedbottom
\setlength{\bibsep}{0pt}
\setlength{\itemsep}{0pt}
\setlength{\parskip}{0pt}
\bibliography{references_clean}

\begin{thebibliography}{}
\makeatletter
\relax
\def\mn@urlcharsother{\let\do\@makeother \do\$\do\&\do\#\do\^\do\_\do\%\do\~}
\def\mn@doi{\begingroup\mn@urlcharsother \@ifnextchar [ {\mn@doi@}
  {\mn@doi@[]}}
\def\mn@doi@[#1]#2{\def\@tempa{#1}\ifx\@tempa\@empty \href
  {http://dx.doi.org/#2} {doi:#2}\else \href {http://dx.doi.org/#2} {#1}\fi
  \endgroup}
\def\mn@eprint#1#2{\mn@eprint@#1:#2::\@nil}
\def\mn@eprint@arXiv#1{\href {http://arxiv.org/abs/#1} {{\tt arXiv:#1}}}
\def\mn@eprint@dblp#1{\href {http://dblp.uni-trier.de/rec/bibtex/#1.xml}
  {dblp:#1}}
\def\mn@eprint@#1:#2:#3:#4\@nil{\def\@tempa {#1}\def\@tempb {#2}\def\@tempc
  {#3}\ifx \@tempc \@empty \let \@tempc \@tempb \let \@tempb \@tempa \fi \ifx
  \@tempb \@empty \def\@tempb {arXiv}\fi \@ifundefined
  {mn@eprint@\@tempb}{\@tempb:\@tempc}{\expandafter \expandafter \csname
  mn@eprint@\@tempb\endcsname \expandafter{\@tempc}}}

\bibitem[\protect\citeauthoryear{Kahn \& Woltjer}{Kahn \&
  Woltjer}{1959}]{kahnIntergalacticMatterGalaxy1959}
Kahn F.~D.,  Woltjer L.,  1959. Intergalactic {{Matter}} and the {{Galaxy}}.,
  \mn@doi [Astrophys. J.] {10.1086/146762}, \href
  {https://ui.adsabs.harvard.edu/abs/doi:10.1086/146762} {130, 705}

\bibitem[\protect\citeauthoryear{{van der Marel} et~al.,}{{van der Marel}
  et~al.}{2012}]{vandermarelM31VelocityVector2012}
{van der Marel} R.~P.,  et~al., 2012. The {{M31 Velocity Vector}}. {{II}}.
  {{Radial Orbit}} toward the {{Milky Way}} and {{Implied Local Group Mass}},
  \mn@doi [Astrophys. J.] {10.1088/0004-637X/753/1/8}, \href
  {https://ui.adsabs.harvard.edu/abs/doi:10.1088/0004-637X/753/1/8} {753, 8}

\bibitem[\protect\citeauthoryear{Li \& White}{Li \&
  White}{2008}]{liMassesLocalGroup2008}
Li Y.-S.,  White S. D.~M.,  2008. Masses for the {{Local Group}} and the
  {{Milky Way}}, \mn@doi [Mon. Not. R. Astron. Soc.]
  {10.1111/j.1365-2966.2007.12748.x}, \href
  {https://ui.adsabs.harvard.edu/abs/doi:10.1111/j.1365-2966.2007.12748.x}
  {384, 1459}

\bibitem[\protect\citeauthoryear{{Lynden-Bell}}{{Lynden-Bell}}{1981}]{lynden-bellDynamicalAgeLocal1981}
{Lynden-Bell} D.,  1981. The Dynamical Age of the Local Group of Galaxies,
  Obs., 101, 111

\bibitem[\protect\citeauthoryear{Giraud}{Giraud}{1986}]{giraudPerturbationNearbyExtragalactic1986}
Giraud E.,  1986. Perturbation of the Nearby Extragalactic Velocity Field by
  the {{Local Group}}, Astron. Astrophys., 170, 1

\bibitem[\protect\citeauthoryear{Sandage}{Sandage}{1986}]{sandageRedshiftDistanceRelationIX1986}
Sandage A.,  1986. The {{Redshift-Distance Relation}}. {{IX}}. {{Perturbation}}
  of the {{Very Nearby Velocity Field}} by the {{Mass}} of the {{Local Group}},
  \mn@doi [Astrophys. J.] {10.1086/164387}, \href
  {https://ui.adsabs.harvard.edu/abs/doi:10.1086/164387} {307, 1}

\bibitem[\protect\citeauthoryear{Karachentsev et~al.,}{Karachentsev
  et~al.}{2002}]{karachentsevVeryLocalHubble2002}
Karachentsev I.~D.,  et~al., 2002. The Very Local {{Hubble}} Flow, \mn@doi
  [Astron. Astrophys.] {10.1051/0004-6361:20020649}, \href
  {https://ui.adsabs.harvard.edu/abs/doi:10.1051/0004-6361:20020649} {389, 812}

\bibitem[\protect\citeauthoryear{Pe{\~n}arrubia, Ma, Walker  \&
  McConnachie}{Pe{\~n}arrubia et~al.}{2014}]{penarrubiaDynamicalModelLocal2014}
Pe{\~n}arrubia J.,  Ma Y.-Z.,  Walker M.~G.,   McConnachie A.,  2014. A
  Dynamical Model of the Local Cosmic Expansion, \mn@doi [Mon. Not. R. Astron.
  Soc.] {10.1093/mnras/stu879}, \href
  {https://ui.adsabs.harvard.edu/abs/doi:10.1093/mnras/stu879} {443, 2204}

\bibitem[\protect\citeauthoryear{Wempe et~al.,}{Wempe
  et~al.}{2024}]{wempeConstrainedCosmologicalSimulations2024}
Wempe E.,  et~al., 2024. Constrained Cosmological Simulations of the {{Local
  Group}} Using {{Bayesian}} Hierarchical Field-Level Inference, \mn@doi
  [Astron. Astrophys.] {10.1051/0004-6361/202450975}, \href
  {https://ui.adsabs.harvard.edu/abs/doi:10.1051/0004-6361/202450975} {691,
  A348}

\bibitem[\protect\citeauthoryear{Jasche \& Wandelt}{Jasche \&
  Wandelt}{2013}]{jascheBayesianPhysicalReconstruction2013}
Jasche J.,  Wandelt B.~D.,  2013. Bayesian Physical Reconstruction of Initial
  Conditions from Large-Scale Structure Surveys, \mn@doi [Mon. Not. R. Astron.
  Soc.] {10.1093/mnras/stt449}, \href
  {https://ui.adsabs.harvard.edu/abs/doi:10.1093/mnras/stt449} {432, 894}

\bibitem[\protect\citeauthoryear{Jasche \& Lavaux}{Jasche \&
  Lavaux}{2019}]{jaschePhysicalBayesianModelling2019}
Jasche J.,  Lavaux G.,  2019. Physical {{Bayesian}} Modelling of the Non-Linear
  Matter Distribution: New Insights into the {{Nearby Universe}}, \mn@doi
  [Astron. Astrophys.] {10.1051/0004-6361/201833710}, \href
  {https://ui.adsabs.harvard.edu/abs/doi:10.1051/0004-6361/201833710} {625,
  A64}

\bibitem[\protect\citeauthoryear{Wempe, Helmi, White, Jasche  \& Lavaux}{Wempe
  et~al.}{2025}]{wempeEffectEnvironmentMass2025}
Wempe E.,  Helmi A.,  White S. D.~M.,  Jasche J.,   Lavaux G.,  2025. The
  Effect of Environment on the Mass Assembly History of the {{Milky Way}} and
  {{M31}}, \mn@doi [Astron. Astrophys.] {10.1051/0004-6361/202553744}, \href
  {https://ui.adsabs.harvard.edu/abs/doi:10.1051/0004-6361/202553744} {701,
  A178}

\bibitem[\protect\citeauthoryear{Tully et~al.,}{Tully
  et~al.}{2008}]{tullyOurPeculiarMotion2008}
Tully R.~B.,  et~al., 2008. Our {{Peculiar Motion Away}} from the {{Local
  Void}}, \mn@doi [Astrophys. J.] {10.1086/527428}, \href
  {https://ui.adsabs.harvard.edu/abs/doi:10.1086/527428} {676, 184}

\bibitem[\protect\citeauthoryear{McCall}{McCall}{2014}]{mccallCouncilGiants2014}
McCall M.~L.,  2014. A {{Council}} of {{Giants}}, \mn@doi [Mon. Not. R. Astron.
  Soc.] {10.1093/mnras/stu199}, \href
  {https://ui.adsabs.harvard.edu/abs/doi:10.1093/mnras/stu199} {440, 405}

\bibitem[\protect\citeauthoryear{Tully \& Fisher}{Tully \&
  Fisher}{1987}]{tullyAtlasNearbyGalaxies1987}
Tully R.~B.,  Fisher J.~R.,  1987. Atlas of {{Nearby Galaxies}}

\bibitem[\protect\citeauthoryear{Peebles, Phelps, Shaya  \& Tully}{Peebles
  et~al.}{2001}]{peeblesRadialTransverseVelocities2001}
Peebles P. J.~E.,  Phelps S.~D.,  Shaya E.~J.,   Tully R.~B.,  2001. Radial and
  {{Transverse Velocities}} of {{Nearby Galaxies}}, \mn@doi [Astrophys. J.]
  {10.1086/321326}, \href
  {https://ui.adsabs.harvard.edu/abs/doi:10.1086/321326} {554, 104}

\bibitem[\protect\citeauthoryear{Peebles \& Nusser}{Peebles \&
  Nusser}{2010}]{peeblesNearbyGalaxiesPointers2010}
Peebles P. J.~E.,  Nusser A.,  2010. Nearby Galaxies as Pointers to a Better
  Theory of Cosmic Evolution, \mn@doi [Nature] {10.1038/nature09101}, \href
  {https://ui.adsabs.harvard.edu/abs/doi:10.1038/nature09101} {465, 565}

\bibitem[\protect\citeauthoryear{Tully, Courtois  \& Sorce}{Tully
  et~al.}{2016}]{tullyCosmicflows32016}
Tully R.~B.,  Courtois H.~M.,   Sorce J.~G.,  2016. Cosmicflows-3, \mn@doi
  [Astron. J.] {10.3847/0004-6256/152/2/50}, \href
  {https://ui.adsabs.harvard.edu/abs/doi:10.3847/0004-6256/152/2/50} {152, 50}

\bibitem[\protect\citeauthoryear{Neuzil, Mansfield  \& Kravtsov}{Neuzil
  et~al.}{2020}]{neuzilSheetGiantsUnusual2020}
Neuzil M.~K.,  Mansfield P.,   Kravtsov A.~V.,  2020. The {{Sheet}} of
  {{Giants}}: {{Unusual}} Properties of the {{Milky Way}}'s Immediate
  Neighbourhood, \mn@doi [Mon. Not. R. Astron. Soc.] {10.1093/mnras/staa898},
  \href {https://ui.adsabs.harvard.edu/abs/doi:10.1093/mnras/staa898} {494,
  2600}

\bibitem[\protect\citeauthoryear{Yaaqib, Petersen  \& Pe{\~n}arrubia}{Yaaqib
  et~al.}{2024}]{yaaqibRadialVariationLMCinduced2024}
Yaaqib R.,  Petersen M.~S.,   Pe{\~n}arrubia J.,  2024. The Radial Variation of
  the {{LMC-induced}} Reflex Motion of the {{Milky Way}} Disc Observed in the
  Stellar Halo, \mn@doi [Mon. Not. R. Astron. Soc.] {10.1093/mnras/stae1363},
  \href {https://ui.adsabs.harvard.edu/abs/doi:10.1093/mnras/stae1363} {531,
  3524}

\bibitem[\protect\citeauthoryear{Chandra et~al.,}{Chandra
  et~al.}{2025}]{chandraAllskyKinematicsDistant2025}
Chandra V.,  et~al., 2025. All-Sky {{Kinematics}} of the {{Distant Halo}}:
  {{The Reflex Response}} to the {{LMC}}, \mn@doi [Astrophys. J.]
  {10.3847/1538-4357/addab6}, \href
  {https://ui.adsabs.harvard.edu/abs/doi:10.3847/1538-4357/addab6} {988, 156}

\bibitem[\protect\citeauthoryear{Karachentsev et~al.,}{Karachentsev
  et~al.}{2003}]{karachentsevLocalGalaxyFlows2003}
Karachentsev I.~D.,  et~al., 2003. Local Galaxy Flows within 5 {{Mpc}}, \mn@doi
  [Astron. Astrophys.] {10.1051/0004-6361:20021566}, \href
  {https://ui.adsabs.harvard.edu/abs/doi:10.1051/0004-6361:20021566} {398, 479}

\bibitem[\protect\citeauthoryear{{Planck Collaboration} et~al.,}{{Planck
  Collaboration} et~al.}{2020}]{planckcollaborationPlanck2018Results2020}
{Planck Collaboration} et~al., 2020. Planck 2018 Results: {{VI}}.
  {{Cosmological}} Parameters, \mn@doi [Astron. Astrophys.]
  {10.1051/0004-6361/201833910}, \href
  {https://ui.adsabs.harvard.edu/abs/doi:10.1051/0004-6361/201833910} {641, A6}

\bibitem[\protect\citeauthoryear{Springel, Pakmor, Zier  \& Reinecke}{Springel
  et~al.}{2021}]{springelSimulatingCosmicStructure2021}
Springel V.,  Pakmor R.,  Zier O.,   Reinecke M.,  2021. Simulating Cosmic
  Structure Formation with the {\textsc{Gadget}} -4 Code, \mn@doi [Mon. Not. R.
  Astron. Soc.] {10.1093/mnras/stab1855}, \href
  {https://ui.adsabs.harvard.edu/abs/doi:10.1093/mnras/stab1855} {506, 2871}

\bibitem[\protect\citeauthoryear{Powell \& Abel}{Powell \&
  Abel}{2015}]{powellExactGeneralRemeshing2015}
Powell D.,  Abel T.,  2015. An Exact General Remeshing Scheme Applied to
  Physically Conservative Voxelization, \mn@doi [J. Comput. Phys.]
  {10.1016/j.jcp.2015.05.022}, \href
  {https://ui.adsabs.harvard.edu/abs/doi:10.1016/j.jcp.2015.05.022} {297, 340}

\bibitem[\protect\citeauthoryear{Harris et~al.,}{Harris
  et~al.}{2020}]{harris2020array}
Harris C.~R.,  et~al., 2020. Array Programming with {{NumPy}}, \mn@doi [Nature]
  {10.1038/s41586-020-2649-2}, \href
  {https://ui.adsabs.harvard.edu/abs/doi:10.1038/s41586-020-2649-2} {585, 357}

\bibitem[\protect\citeauthoryear{Virtanen et~al.,}{Virtanen
  et~al.}{2020}]{2020SciPy-NMeth}
Virtanen P.,  et~al., 2020. {{SciPy}} 1.0: {{Fundamental}} Algorithms for
  Scientific Computing in Python, \mn@doi [Nat. Methods]
  {10.1038/s41592-019-0686-2}, \href
  {https://ui.adsabs.harvard.edu/abs/doi:10.1038/s41592-019-0686-2} {17, 261}

\bibitem[\protect\citeauthoryear{Hunter}{Hunter}{2007}]{Hunter:2007}
Hunter J.~D.,  2007. Matplotlib: {{A 2D}} Graphics Environment, \mn@doi
  [Comput.. Sci. Eng.] {10.1109/MCSE.2007.55}, \href
  {https://ui.adsabs.harvard.edu/abs/doi:10.1109/MCSE.2007.55} {9, 90}

\bibitem[\protect\citeauthoryear{{Astropy Collaboration} et~al.,}{{Astropy
  Collaboration} et~al.}{2018}]{astropycollaborationAstropyProjectBuilding2018}
{Astropy Collaboration} et~al., 2018. The {{Astropy Project}}: {{Building}} an
  {{Open-science Project}} and {{Status}} of the v2.0 {{Core Package}}, \mn@doi
  [Astron. J.] {10.3847/1538-3881/aabc4f}, \href
  {https://ui.adsabs.harvard.edu/abs/doi:10.3847/1538-3881/aabc4f} {156, 123}

\bibitem[\protect\citeauthoryear{{Astropy Collaboration} et~al.,}{{Astropy
  Collaboration}
  et~al.}{2022}]{astropycollaborationAstropyProjectSustaining2022}
{Astropy Collaboration} et~al., 2022. The {{Astropy Project}}: {{Sustaining}}
  and {{Growing}} a {{Community-oriented Open-source Project}} and the {{Latest
  Major Release}} (v5.0) of the {{Core Package}}, \mn@doi [Astrophys. J.]
  {10.3847/1538-4357/ac7c74}, \href
  {https://ui.adsabs.harvard.edu/abs/doi:10.3847/1538-4357/ac7c74} {935, 167}

\bibitem[\protect\citeauthoryear{Coelho}{Coelho}{2017}]{coelhoJugSoftwareParallel2017}
Coelho L.~P.,  2017. Jug: {{Software}} for {{Parallel Reproducible
  Computation}} in {{Python}}, \mn@doi [J. Open Res. Softw.]
  {10.5334/jors.161}, \href
  {https://ui.adsabs.harvard.edu/abs/doi:10.5334/jors.161} {5}

\makeatother
\end{thebibliography}
\endgroup

\section*{Acknowledgements}

The Gadget simulations were performed on the Freya cluster at the Max Planck Computing and Data Facility. The constrained simulations were enabled by resources provided by the Swedish National Infrastructure for Computing at the PDC Center for High Performance Computing, KTH Royal Institute of Technology, partially funded by the Swedish Research Council (Grant Agreement No. 2018-05973). J.J. and G.L. acknowledge support from the Simons Foundation through the Simons Collaboration on Learning the Universe. This work was made possible by the research project grant Understanding the Dynamic Universe, funded by the Knut and Alice Wallenberg Foundation (Grant No. Dnr KAW 2018.0067). Additionally, J.J. acknowledges financial support from the Swedish Research Council through the project Deciphering the Dynamics of Cosmic Structure (Grant No. 2020-05143). G.L. acknowledges support from the CNRS-IEA Manticore project. This work has been financially supported by a Spinoza Prize from NWO to A.H. (Grant No. SPI 78 411). This work was done in part within the Aquila Consortium (\url{https://www.aquila-consortium.org}). E.W. thanks Akshara Viswanathan for her moral support.

During this work, we made use of various software packages: \borg\ \citem{jascheBayesianPhysicalReconstruction2013,jaschePhysicalBayesianModelling2019}, \textsc{Gadget-4} \citemet{springelSimulatingCosmicStructure2021}, \textsc{r3d} \citemet{powellExactGeneralRemeshing2015}, \textsc{NumPy} \citemet{harris2020array}, \textsc{SciPy} \citemet{2020SciPy-NMeth}, \textsc{Matplotlib} \citemet{Hunter:2007}, \textsc{H5PY}, \textsc{Astropy} \citemet{astropycollaborationAstropyProjectBuilding2018,astropycollaborationAstropyProjectSustaining2022} and \textsc{Jug} \citemet{coelhoJugSoftwareParallel2017}.

\section*{Author contributions}

E.W. performed the simulations, created the figures and wrote the paper. S.D.M.W. supervised the project and was instrumental in the analysis and development of the article. A.H. provided supervision and contributed to the analysis and the paper. G.L. and J.J. assisted with the simulations and provided valuable input throughout the project.

\section*{Competing interests}

The authors declare no competing interests.

\makeatletter
\makeatother

\begin{figure*}
    \centering
    \includegraphics[width=0.7\linewidth]{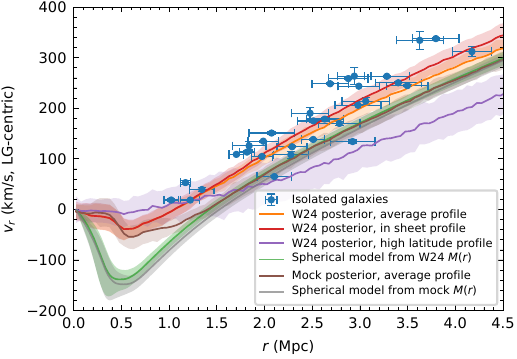}
    \caption{%
    \textbf{Anisotropic local Hubble flow and effects of tracer positions. } 
    Expanded Hubble diagram showing line-of-sight recession velocity relative to the Local Group barycentre as a function of distance. As in  \cref{fig:hubblediagram}, blue points are the 31 isolated galaxies used to constrain the local flow and the orange curve shows the mass-weighted mean radial velocity of all matter in the constrained simulation ensemble, with the bands showing the rms scatter between realizations.
    The red curve shows the mass-weighted velocity-distance relation for material lying in the sheet (defined as $\abs{z}<\SI{1}{Mpc}$). It lies close to (but slightly above) the orange curve which gives the same quantity for all matter. The purple curve gives the corresponding relation for high-latitude material ($\abs{b}>45$~deg) and is very different, particularly at larger distances where the differences in recession velocity reach \SI{\sim100}{km.s^{-1}}. This substantial anisotropy of the infall velocity field is also evident in the flow field shown in \cref{fig:velocityfield} where infall is directed primarily towards the Local Sheet, rather than towards the Local Group. To validate our results in light of the anisotropic sky distribution of our tracer population, we performed a mock inference in which tracer galaxy velocities were replaced by predictions from a spherical model using the mean mass profile $M(r)$ (the green line, identical to \cref{fig:hubblediagram}'s spherical model). The resulting mock ensemble's velocity profile is the brown line, which matches the original ensemble (the orange line) at small radii but approaches the spherical model beyond \SI{1}{Mpc}. The grey line shows a spherical model using the $M(r)$ of the mock inference, which matches that of the original inference. Thus, despite the highly anisotropic distribution of tracers, our inference scheme correctly reproduces $M(r)$ when the underlying mass distribution is, in fact, spherical.}
    \label{fig:hubblediagramwmock}
\end{figure*}
\begin{figure*}
    \centering
    \includegraphics[width=\linewidth]{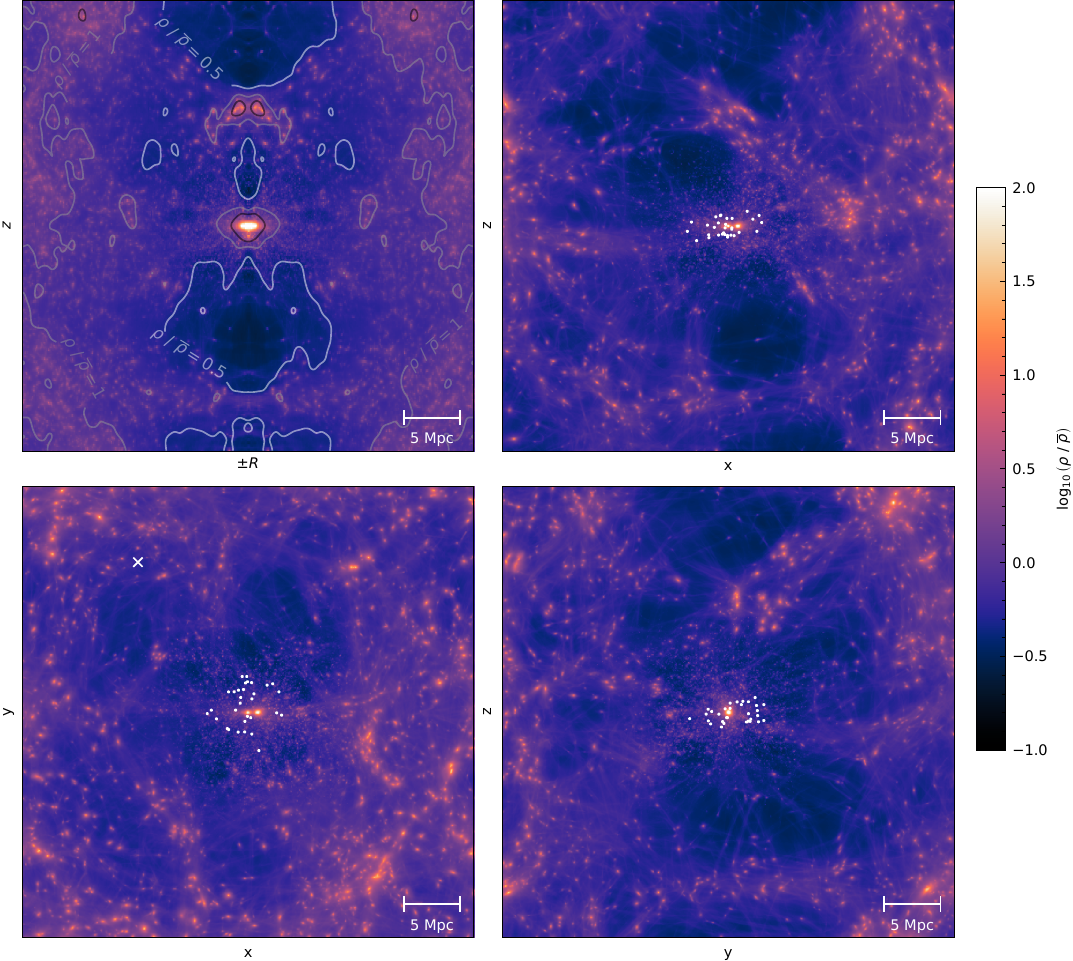}
    \caption{%
    {\textbf{Posterior mean density field in a spherical mock inference.} Posterior mean matter density field (like \cref{fig:sheet}) for the mock ensemble of simulations where all constraints were the same except that the flow tracer velocities were replaced by predictions from a spherical model with the same $M(r)$. This posterior mean field is estimated from 117 simulations homogeneously sampled from the HMC chains. Unlike the results in \cref{fig:sheet}, this mock posterior mean field exhibits no clear preferred plane or sheet-like mass concentration, confirming that the sheet-like structure inferred from the real data is driven by the observed kinematics. Interestingly, the posterior density fields obtained from this mock dataset are not fully isotropic; despite tracer velocities being assigned spherically, a significant asymmetry remains between the $+z$ and $-z$ directions. We suspect this is due to the only constraint that breaks the $\pm z$ symmetry: M31's tangential velocity. This requires an appropriate torque from external matter. A quadrupolar matter distribution with the right orientation is visible about $\SI{8}{Mpc}$ from the centre (e.g., top right panel), suggesting it accommodates the tangential constraint when the Local Sheet is absent. The isotropic expansion of observed tracers in both sets of chains is presumably inconsistent with the in-plane quadrupole needed to generate M31's observed $y$-component proper motion, which is not fully reproduced in either case.}}
    \label{fig:sheetmock}
\end{figure*}
\begin{figure*}
    \centering
    \includegraphics[width=0.7\linewidth]{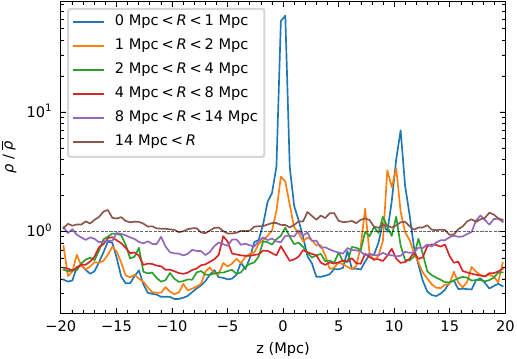}
    \caption{%
    {\textbf{Vertical density profiles for the spherical mock inference.} Azimuthally averaged density as a function of height $z$ within thick cylindrical shells around the MW-M31 barycentre for the mock ensemble in which tracer velocities are replaced by predictions from a spherical model with the same enclosed mass profile $M(r)$ as in the constrained ensemble. Quantitatively, these profiles confirm the visual impression from the previous figure: the mock density profiles exhibit no preferred plane. This further confirms that the sheet-like structure inferred from the real data is due to the velocities of the tracer galaxies, not their flattened spatial distribution. Interestingly, there is still an indication that the region surrounding the Local Group must be generally underdense in order to achieve the mass density profile which has been assumed in this spherical-velocity mock test.}}
    \label{fig:densityringsmock}
\label{LastPage}
\end{figure*}

\end{document}